\documentclass{raa}            
\usepackage{graphicx,times}
\usepackage{natbib}
\usepackage{textcomp}
\usepackage{caption}
\usepackage{gensymb}
\usepackage{amssymb,amsmath}
\usepackage{longtable}
\bibpunct{(}{)}{;}{a}{}{,}
\usepackage[driverfallback=dvipdfm]{hyperref}
\hypersetup{colorlinks = true, linkcolor = green, anchorcolor = red, citecolor = blue, filecolor = red, urlcolor = red}
\usepackage[margin=1in,a4paper]{geometry}  
\usepackage{array,booktabs,ragged2e}
\newcolumntype{R}[1]{>{\RaggedLeft\arraybackslash}p{#1}}

\begin{document}

   \title{Discovery of 2716 hot emission-line stars from LAMOST DR5}

   \volnopage{Vol.0 (20xx) No.0, 000--000}      
   \setcounter{page}{1}          

   \author{ B. Shridharan
      \inst{1}
   \and Blesson Mathew
      \inst{1}
   \and S. Nidhi
      \inst{1}
   \and R. Anusha
      \inst{1}
   \and R. Arun
      \inst{1}
    \and Sreeja S. Kartha
       \inst{1}
    \and Yerra Bharat Kumar 
       \inst{2}
   }

  {
   \institute{ Department of Physics and Electronics, CHRIST (Deemed to be University), Bangalore 560029, India; {\it{shridharan.b@res.christuniversity.in}}
   \\ 
   \and
        Key Laboratory of Optical Astronomy, National Astronomical Observatories, Chinese Academy of Sciences, Beijing 100101, China \\
   }
       \vs\no
   {\small Received~~20xx month day; accepted~~20xx~~month day}}

\abstract{We present a catalog of 3339 hot emission-line stars (ELS) identified from 451,695 O, B, and A type spectra, provided by the LAMOST DR5 release. We developed an automated python routine that identified 5437 spectra having a peak between 6561 and 6568~\AA. False detections and bad spectra were removed, leaving 4138 good emission-line spectra of 3339 unique ELS. We re-estimated the spectral type of 3307 spectra as the LAMOST Stellar Parameter Pipeline (LASP) did not provide accurate spectral types for these emission-line spectra. As Herbig Ae/Be stars show higher excess in near-infrared and mid-infrared wavelengths than Classical Ae/Be stars, we used 2MASS and WISE photometry to distinguish them. Finally, we report 1089 Classical Be, 233 Classical Ae, and 56 Herbig Ae/Be stars identified from the LAMOST DR5. In addition, 928 B[em]/A[em] stars and 240 CAe/CBe potential candidates are identified. From our sample of 3339 hot emission-line stars, 2716 ELS identified in this work do not have any record in the SIMBAD database and they can be considered as new detections. Identification of such a large homogeneous set of emission-line spectra will help the community to study the emission phenomenon in detail without worrying about the inherent biases when compiling from various sources.
}


   \authorrunning{Shridharan et al.}            
   \titlerunning{Hot emission-line stars in LAMOST DR5}  

   \maketitle
\section{Introduction}           
\label{sect:intro}

Stars with bright emission lines in the spectrum indicating unusual activities in the stellar atmosphere has become a subject of much astrophysical research. Objects with such diverse spectra showing Balmer emission along with metallic emission lines are generally known as Emission-line stars (ELS). These lines arising from the circumstellar disk/envelope of the star are formed due to the recombination process. Based on the evolutionary stage, spectral type, and mass, ELS can be classified into various categories such as Young Stellar Objects (YSOs), Oe/Be/Ae stars, Wolf-Rayet (WR) stars, Supergiants, Planetary Nebulae (PNe), Luminous Blue Variables (LBVs), Mira stars, Flare stars, and Symbiotic stars \citep{2007ASPC..362..260K}. Hence, the study of ELS can provide insight towards the line formation regions and various evolutionary phases of stars.

In recent years, research on ELS \citep{2020A&A...638A..21V, 2019ApJS..240...21A, 2013AJ....145..126H} has surged due to the availability of photometric surveys (\textit{Gaia}, 2MASS, IPHAS, etc.) and large scale spectroscopic surveys (LAMOST, SDSS, APOGEE). \cite{1949ApJ...110..387M} were the first to undertake a survey on ELS in which spectra of early-type stars with bright hydrogen lines were detected. \cite{1970MNRAS.149..405W} compiled a catalogue of 5,326 early-type ELS in our Galaxy. Later, \cite{1973MNRAS.161..145A} observed nearly 250 early-type ELS in the near-infrared (NIR) wavelengths and proposed a mechanism wherein the excess flux in NIR (IR excess) arises due to the thermal radiation from the dust in the circumstellar disk and the electron \textit{bremsstrahlung} originating from a shell of ionized gas. Following these works,  \cite{stephenson1977new} identified 455 new bright H$\alpha$ stars in our galaxy using objective-prism plates. By the end of 20th century, various surveys increased the count of ELS considerably. One of the most extensive ELS survey in H{$\alpha$} regime was performed by \cite{1999A&AS..134..255K}, in which they provided a catalog of 4174 H$\alpha$ ELS in the Northern Milky Way within the latitude range of -10$^{\circ}$ $< b <$ 10$^{\circ}$. Following \cite{1999A&AS..134..255K}, a much larger and deeper survey was performed and is now known as, The INT Photometric H$\alpha$ Survey of the Northern Galactic Plane (IPHAS; \citealp{2005MNRAS.362..753D}). It covered the Milky Way in \textit{H$\alpha$}, \textit{Sloan} \textit{r'} and \textit{i'} bands within a magnitude range of 13 $\leq$~\textit{r'} $\leq$~19.5 mag. Following the initial data release from IPHAS \citep{gonzalez2008initial}, a series of studies on different types of ELS were published. \cite{2008MNRAS.384.1277W} presented the preliminary catalogue of 4853 H$\alpha$ emission sources from IPHAS. Various ELS classes including cataclysmic variables \citep{2007MNRAS.382.1158W}, symbiotic stars \citep{2008A&A...480..409C,2010A&A...509A..41C,2014A&A...567A..49R}, YSOs \citep{vink2008iphas,2011MNRAS.415..103B} and classical Be stars \citep{2013MNRAS.430.2169R} were identified from the survey. Similarly,  \cite{2015MNRAS.453.1026K} has studied the accretion rates of 235 Classical T Tauri star (CTTs) candidates in the Lagoon Nebula using the VST Photometric H$\alpha$ Survey of the Southern Galactic Plane and Bulge (VPHAS+; \citealp{drew2014vst}). There are quite a few studies which searched for ELS in open clusters. For example, \citet{mathew2008phenomenon} identified 157 ELS from the slitless spectroscopic survey of 207 open clusters in the Galaxy. \cite{2004AJ....127.1117R} and \cite{2014A&A...570A..30P} presented H$\alpha$ ELS surveys in various molecular clouds using the wide field-objective prism films. Recently, \cite{2020A&A...638A..21V} identified new ELS candidates using machine learning techniques on data from \textit{Gaia} DR2, 2MASS, WISE and IPHAS or VPHAS+ surveys. 

The Large Sky Area Multi-Object fiber Spectroscopic Telescope (LAMOST) surveys have accumulated millions of spectra over a period of 5-6 years. Even though a huge spectral database from LAMOST is made available to the public, very few works have identified and characterized ELS spectra till date. \cite{2015RAA....15.1325L} identified 192 Classical Be (CBe) star candidates from LAMOST DR1, which increased the then known sample of CBe stars by about 8\%. Similarly, a study on ELS and H$\alpha$ line profiles from LAMOST DR2 were conducted by \cite{2016RAA....16..138H}, in which 10,436 early-type ELS were identified.  \cite{2017ApJS..232...16Y} have studied Mira Variable Stars identified from LAMOST DR4 and \cite{2018ApJ...863...70L} identified 6 new Oe stars from LAMOST DR5. 


In this work, we make use of the extensive data available to us from LAMOST DR5 release to identify early-type ELS and mainly classify them into CBe stars and Herbig Ae/Be stars (HAeBe). A CBe star is a fast rotating B-type non-super giant  with an equatorial decretion disk \citep{2003PASP..115.1153P}. The disk is transient in nature identified through the change in H$\alpha$ emission strength over a timescale of years to decades \citep{2003A&A...411..167S}. The formation of decretion disk in CBe stars is known as `Be phenomenon’. The episodic occurrence of mass loss which results in the formation of a decretion disk is still an open question. The emission strength of H$\alpha$ and other hydrogen and metallic emission lines were analyzed to understand the `Be phenomenon' in CBe stars \cite[for a review]{2013A&ARv..21...69R}. Most CBe stars span the spectral range of B0 - B9, even though a few late O \citep{2018ApJ...863...70L} and early A type stars \citep{jaschek1986type,2021MNRAS.501.5927A} are included in this class. Also, CBe stars are defined to belong in the luminosity class III to V \citep{1982IAUS...98..261J}. In contrast to CBe stars, pre-main sequence (PMS) stars are very young objects that have not started hydrogen burning. The low-mass (0.1 to 2 M$_{\odot}$) PMS stars are known as T Tauri stars \citep{1945ApJ...102..168J}, whereas those in the intermediate-mass range (2 to 8 M$_{\odot}$) are called HAeBe stars \citep{1960ApJS....4..337H,1998ARA&A..36..233W}. HAeBe and T Tauri stars share common characteristics such as the circumstellar accretion disk, IR excess, Balmer emission lines and metallic emission lines \citep{1992ApJ...397..613H}. For the present study, we discuss the ELS belonging to O, B and A spectral types, whereas those belonging to later spectral types will be discussed in a follow-up study (Edwin et al. in prep).

We downloaded spectra of O, B, and A type stars from LAMOST DR5 and developed a python routine to identify spectra with H$\alpha$ in emission. To re-estimate the spectral types of ELS spectra, we performed a visually aided semi-automated template matching process using MILES spectral library. Further, we made use of available photometric data from missions/surveys such as 2MASS, WISE, \textit{Gaia} EDR3 and APASS to sub-classify our sample into CBe stars and HAeBe stars. We categorized the identified ELS spectra into various classes, as explained in subsection \ref{subsec:class}.

The structure of the paper is as follows. Section \ref{sect:Data} provides a brief introduction of the LAMOST survey and explains the data collection along with the spectra normalization procedure. In Section \ref{sect:Analysis}, we discuss the ELS identification and classification procedure along with the explanation of spectral-type re-estimation technique. We have also detailed the naming convention employed in this work. In Section \ref{sect:result}, we plot infrared color-color diagram (CCDm) and \textit{Gaia} color-magnitude diagram (CMD), followed by discussion of spectral features seen in different classes identified in this work. We cross-matched our list with the SIMBAD database to verify our classifications. The major results of our work are summarized in Section \ref{sect:conclusion}. Description of our catalog is given in Appendix \ref{appdx:A}, automated forbidden line identification criteria is given in Appendix \ref{appdx:B} and necessary information for a representative sample of ELS identified are given in Appendix \ref{appdx:C}.

\section{Data Collection}
\label{sect:Data}

\subsection{About LAMOST Data Release 5}
The LAMOST is a reflecting Schmidt telescope, with an effective aperture of 3.6~m - 4.9~m and a wide Field of View (FoV) of 5$^\circ$ \citep{2012RAA....12.1197C, 1996ApOpt..35.5155W}. It is also known as Guo Shoujing Telescope, maintained by Xinglong station of the National Astronomical Observatories, Chinese Academy of Sciences. It is equipped with 4000 optical fibers in its focal plane with 16 low-resolution spectrographs and 32 Charge-Coupled Devices (CCD) \citep{zhao2012lamost}. It is designed with three major components: the correcting mirror \textit{Ma}, the primary mirror \textit{Mb} and a focal surface \citep{2012RAA....12.1197C}. Objects with \textit{r}-band magnitude up to $\sim$20 mag are observed by LAMOST in the wavelength range of 3690 – 9000~\AA~at spectral resolution of \textit{R}$\sim$1800 \citep{2012RAA....12.1197C}. The Galactic and Extra Galactic surveys conducted by LAMOST are classified into two major projects: LAMOST Experiment for Galactic Understanding and Exploration (LEGUE; \citealp{2012RAA....12..735D}) and LAMOST Extra GAlactic Survey (LEGAS; \citealp{zhao2012lamost}).

 
\subsection{Data Collection and Spectral Normalization}

During the five-year regular survey, the raw data was reduced by LAMOST two-dimensional (2D) pipeline \citep{luo2015first}, which used procedures similar to those of SDSS spectro2d pipeline \citep{stoughton2002sloan}. A spectrum of a light beam from fiber is imaged onto the 32 cameras using $4k${$\times$}$4k$ CCD chips \citep{wei1996new}. The data from each CCD chip is separately reduced and then combined with other exposures to improve signal. Bias and dark frames are subtracted from each raw image. Flat-fields are then traced and extracted for each fiber. Wavelength is calibrated using the arc lamp spectra and slightly adjusted to match the known positions of certain sky lines, which is then corrected to heliocentric frame. Further, the LAMOST one-dimensional (1D) pipeline extract and classify spectra into 4 categories (\textit{star}, \textit{galaxy}, \textit{QSO} and \textit{unknown}) where radial velocities, redshifts and other stellar parameters are calculated wherever possible \citep{zhao2012lamost, 2012RAA....12.1197C}. 

From 8,183,160 spectra classified as `\textit{star}' in LAMOST DR5, we selected 451,695 spectra belonging to O, B and A spectral type as given by LAMOST pipeline. LAMOST spectra with low Signal-to-Noise Ratio (SNR) may affect the selection and analysis of H$\alpha$ emission lines. In a study of identifying H$\alpha$ emission spectra from LAMOST DR2 by \cite{2016RAA....16..138H}, spectra with SNR of SDSS \textit{r} band less than 10 ($SNR_r<10$) were removed. Further, the automated routine to find emission lines in LAMOST spectra does not yield good accuracy if we reduce SNR$_r$ band less than 10 ($SNR_r<10$). We observed that many noise peaks were reported as emission peaks. Hence, we adopted the criterion of $SNR_r>10$ in order to remove the noisy spectra from our sample.

The reduced and calibrated spectra were downloaded from LAMOST DR5 website\footnote{dr5.lamost.org}. The data array of a LAMOST fits file is as follows. The first frame contains the flux information, and the second frame contains the ‘inverse variance’ (1/{$\sigma^{2}$}, where $\sigma$ is the uncertainty). The third frame stores wavelength in \AA~where as fourth and fifth frames contain `andmask' and `ormask' data. Further information about the fits description and data structure is accessible online \footnote{http://dr5.lamost.org/v3/doc/data-production-description}. To normalize the spectra, we used \texttt{laspec}, a LAMOST spectral kit developed by Bo Zhang \citep{2020ApJS..246....9Z}. This python-based package is open-source and available in GitHub\footnote{https://github.com/hypergravity/laspec}. The entire spectrum is first smoothed using a smoothing spline and the pixels lying away from the 1.5$\sigma$ threshold is removed. The remaining pixels are smoothed to obtain the pseudo-continuum, which is then used to obtain normalized spectrum. Refer to Section 2.1 of \cite{2020ApJS..246....9Z} for a detailed explanation regarding pre-processing of LAMOST spectra.

\section{Analysis}
\label{sect:Analysis}
\subsection{Identification of Emission-Line Stars from LAMOST DR5}

In this section, we explain the method to identify the ELS with H$\alpha$ in emission in the Galaxy from the large data-set of LAMOST DR5.  For the purpose of identifying ELS with H$\alpha$ in emission, we developed a python code based on \texttt{find\_peaks} module in \texttt{scipy.signal} package \citep{2020SciPy-NMeth}. The important parameter in $\texttt{scipy.signal.find\_peaks}$ is  ``Width", which corresponds to the full width at half maximum (FWHM) of the emission profile. Spikes in the spectrum caused by instrumental defects will be very narrow and can be mistaken for a narrow emission line. To remove such lines, we used a width cutoff of 3 sampling points. This reduces the possibility of false emission (caused by instrumental defects or noise) near H$\alpha$ being reported as real emission in the spectrum. We consider H$\alpha$ emission to be true only if Full Width at Half Maximum (FWHM) is at least 3 sampling points wide. Even though it affects the completeness of ELS from DR5, probability of reporting false emission is reduced compared to selection criteria used in \cite{2016RAA....16..138H}. The scope of this work is to provide a comprehensive catalog of stars with H$\alpha$ in emission which justifies the use of our stringent selection criteria. Out of 451,695 O, B, and A  spectra from LAMOST DR5, we identified 5,437 spectra with H$\alpha$ in emission. Through a visual check, those without any photospheric features and broken spectra were removed, resulting in 4138 good quality spectra with H$\alpha$ in emission from LAMOST DR5. The selection scheme and the classification procedure we followed to identify 4138 emission-line spectra from LAMOST DR5 is represented in the form of a flowchart in Figure \ref{fig:flochart}. It should be noted that if FWHM criterion is relaxed to a lower value, say to 1 sampling point, the number of ELS found would increase drastically but it may consist of many false detections.  

\begin{figure}
    \centering
    \includegraphics[width=1.1\textwidth]{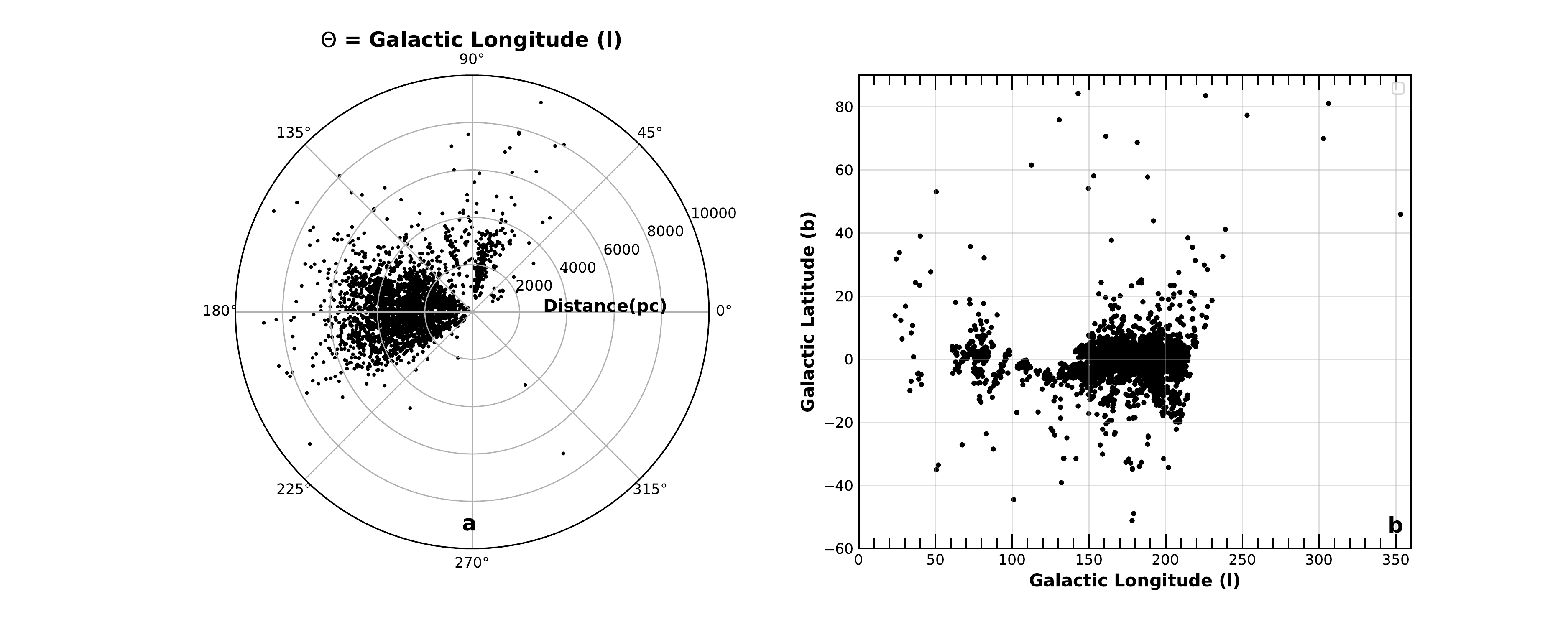}
    \caption{The polar and spatial distribution of 3339 ELS in the Galaxy identified from LAMOST DR5 is shown. Fig 1a (left) shows the polar distribution of ELS in the Galactic Longitude ($l$) vs Distance ($d$) plane. Distances are taken from \citet{2021AJ....161..147B}. The ELS are concentrated in the Galactic anti-center direction due to the observational strategy adopted by LAMOST \citep{2016RAA....16..138H}. Fig 1b (right) represents the spatial distribution of ELS in Galactic Latitude ($b$) vs Galactic  Longitude ($l$) plane. It can be seen that most of the ELS are found to be within $150^\circ\leq l \leq200^\circ$ and $|b|\leq 20^\circ$.}
    \label{fig:lvb}
\end{figure}
 
  From Figure \ref{fig:lvb}, it can be seen that most of the ELS belong to the Galactic anti-centre direction while a few objects are within $60\degree \leq l \leq 90\degree$. \cite{2015MNRAS.448..855Y} reported that the LAMOST Spectroscopic Survey of the Galactic Anti-centre (LSS-GAC) survey initially aimed to observe the Galactic thin/thick disks and halo for a contiguous sky area which is centred on the Galactic anti-centre direction (150$\degree\leq$ l $\leq210\degree$). Also, for large distances, the inverse relation between \textit{Gaia} EDR3 \citep{2021A&A...649A...2L} parallax and distance fails. This was resolved by \cite{2021AJ....161..147B} using a probabilistic approach to determine the distance from \textit{Gaia} EDR3 parallax. The distance range observed here shows that the sample of ELS observed is spread over 10 kpc with about 90\% of the objects within 5.5 kpc. 53 objects are present in the high galactic latitude region ($|b|\geq 40^\circ$) which may be of further interest to readers, but is out of scope of this work.

\subsection{Spectral Type Re-estimation}

The spectral type of each LAMOST DR5 spectrum is estimated using LAMOST stellar parameter pipeline (LASP; \citealp{wu2011automatic}). The raw CCD images are processed through the 2-dimensional (2D) reduction pipeline \citep{2012RAA....12.1197C} and the output is in the form of 1-dimensional (1D) spectrum for each star. The spectral type of the star is estimated through template matching with standard spectral templates and line recognition algorithms \citep{luo2015first}. A sample of 183 stellar spectra from LAMOST pilot survey and LAMOST DR1 were selected as standard spectral templates \citep{wei2014construction,kong2019lamost}. For classifying O and B type stars, LAMOST pipeline has used only O, B, B6, and B9 stellar templates. The presence of the veiled continuum and emission lines formed outside the photosphere complicate the spectral classification in early-type stars \citep{2004AJ....127.1682H}. For this reason, automated spectral type estimation for ELS may give erratic results. This was validated by our finding that, despite eliminating bad spectra from our sample, spectra classified as A-type by LASP show He{\sc i} absorption lines. Hence we need to re-estimate the spectral types for all the spectra. Thus, one has to resort back to semi-automated template matching technique where the user needs to select the best fit considering various spectral lines. In order to distinguish O, B and A type spectra, we can use the profile of Balmer absorption lines. Since we are dealing with emission-line spectra, we cannot use H$\alpha$, H$\beta$ and even H$\gamma$ to some extent because the emission mechanism can fill in these absorption lines. Hence as a first check, we matched the H$\delta$ and H$\epsilon$ profiles to estimate the spectral type coarsely. To estimate the spectral sub-class, we used absorption lines such as He{\sc ii}, He{\sc i} and Mg {\sc ii}. Theoretically, one expects ionized He{\sc ii} lines in O-type stars. The He{\sc ii} lines fade as we move towards B-type stars where He{\sc i} absorption lines are stronger and further He{\sc i} lines slowly disappear in early A-type stars \citep{2009ssc..book.....G}. Also, since CBe stars are fast rotators (200-300 $kms^{-1}$), the depth or profile shape vary depending on the vsin\textit{i} parameter. This will in turn affect the spectral type re-estimation process if we focus on the depth of lines \citep{1980ApJ...242..171S}. Keeping this in mind, we matched the wing profile of H$\epsilon$, H$\delta$, He{\sc i} and Mg{\sc ii} lines rather than the depth of the line. However, estimating accurate spectral type for CBe stars using low-resolution spectra is difficult and one can expect up to an error of 2 sub-classes in B-type stars. To see the comparison between LAMOST candidate spectra and MILES template spectra, please refer to Figure 2 of \cite{2021MNRAS.501.5927A}. 

As explained above, we performed spectral type re-estimation using semi-automated template matching method. For this, we used the MILES stellar spectral library \citep{2006MNRAS.371..703S} since it has similar resolution (R$\sim$2000) to LAMOST DR5 spectra (R$\sim$1800).  The homogeneously calibrated 985 spectra in MILES spectral library were obtained from the 2.5~m Isaac Newton Telescope (INT), covering a range of 3525-7500~\AA~\citep{2011A&A...532A..95F}. The template stars cover a wide parametric space, i.e. T{$_{eff}$} $=$ 3000 -- 40,000 K, log ({$g$}) = 0.2 -- 5.5. Out of 985 MILES templates, 79 of them are in the spectral range B0 -- A9 covering luminosity classes {\sc III} -- {\sc V}. Rest-frame correction to LAMOST DR5 spectrum were made using the redshift value provided by LASP. We developed a python code, which for a given spectra identifies 3 best-fitting templates (based on the chi-square value. Lower the chi\-square value, better the fit) after iterating over the entire MILES library. To make sure the He{\sc i} and Mg{\sc ii} absorption lines fit well, we over-plotted each spectra with top 3 template matches and visually identified the best among them. In most cases, the least chi-square valued template fitted the best to the spectra, but in a few cases, second or third least chi-square valued template fitted the higher-order Balmer lines (H$\delta$ and H$\epsilon$), He{\sc i} 4471 \AA~and Mg{\sc ii} 4481 \AA~lines better, which justifies the time-intensive spectral typing procedure. Out of 4138 ELS spectra, we were able to ascertain spectral types to 3307 spectra. At this stage, spectra showing nebular forbidden lines such as [NII], [SII] and [OI] were identified using an automated python routine, which is explained in Appendix \ref{appdx:B}. These spectra are classified as either A[em] or B[em]\footnote{It should be noted that [em] is used to denote spectra with forbidden lines instead of [e] because [e] stars are conventionally defined to have [Fe{\sc ii}] lines \citep{1973A&A....26..443S}} based on re-estimated spectral types. We could not assign a spectral type to 831 spectra because the blue end of the spectra were noisy and they are classified as ``Em” or ``Em[em]” (if forbidden lines are present).

\subsection{Evolved Star Candidates}
\label{evolved}


The evolved and highly luminous stars (post-main sequence/supergiants) with early spectral type are reported to have H$\alpha$ in emission \citep{Rosendhal1973ApJ...186..909R}. The H$\alpha$ emission in evolved stars primarily originates from wind driven mass loss \citep{Weymann1963ARA&A...1...97W, Hutchings1970MNRAS.147..161H}. The individual reddening for each spectrum co-ordinate was obtained using the \texttt{dustmaps} open source package \citep{2018JOSS....3..695M}, which provides probabilistic reddening measurements based on \textit{Gaia} parallax and stellar photometry from 2MASS and PanSTARRS 1. The reddening $E(g-r)$ obtained using \texttt{dustmaps} is converted to $E(B-V)$ applying a general correction of 0.884\footnote{\url{argonaut.skymaps.info/usage}}.
Using these reddening measurements, we calculated the extinction in the $V$ band by applying the relation A$_V$ = 3.1 {$\times$} $E(B-V)$. The extinction in photometric bands of \textit{Gaia} and 2MASS is then determined by adopting the conversion relations from \citet{2019ApJ...877..116W}.

To identify evolved stars present in our sample, two methods, spectroscopic (first) \& photometric (second), are used. First, 42 spectra with luminosity class Ia/b \& II estimated using MILES library are classified as ``Evolved*"\footnote{`*' is used here to denote that they are Evolved star candidates, a convention we have used throughout this work.}. In addition, there can be some evolved stars whose luminosity class is misclassified. Hence as a second criteria, we used bolometric luminosity (L\textsubscript{bol}) limit adopted from \citet{lamers1998improved}. The criteria of $log(L\textsubscript{bol}/ L_{\odot})$ $>$ 4.5 applied to all the remaining stars. We used \textit{Gaia} BP ($G_{BP}$) magnitude and performed necessary bolometric corrections to calculate $log(L\textsubscript{bol}/ L_{\odot})$. The equation for bolometric magnitude using G\textsubscript{BP} is given below, 

\begin{equation}
    M_{bol} = G_{BP} - 5log(d) + 5 + BC - A_{G_{BP}}
\end{equation}

where BC is the bolometric correction taken from \citet{2010A&A...523A..48J} for the re-estimated spectral types. $A_{G_{BP}}$ is calculated using conversion co-efficient taken from \cite{2019ApJ...877..116W}, which is given as $A_{G_{BP}}$ = $1.002 \times A_{V}$. The $(L\textsubscript{bol}/ L_{\odot})$ is calculated using the following equation,

\begin{equation}
    \frac{L\textsubscript{bol}}{L_{\odot}} = \frac{L\textsubscript{o}}{L_{\odot}} \times 10^{-0.4 M_{bol} }
\label{eqn:2}
\end{equation}

where L\textsubscript{o} is the zero-point luminosity equals $3.0128 \times 10^{28}$ W. Using the above equation, we classified 4 additional spectra as ``Evolved*" as they satisfy the photometric criteria. In total, 46 spectra are classified as `Evolved*', of which 26 spectra belong to B2 or earlier spectral type.

\subsection{Estimation of IR excess in ELS}
\label{subsec:sed}
The spectral energy distribution (SED) of main-sequence and PMS ELS show excess flux over the blackbody distribution in the infrared. This is indicative of the presence of a circumstellar disk in ELS \citep{1993A&A...275..527N}.
Infrared continuum excess observed in CBe stars is due to thermal \textit{bremsstrahlung} emission from the circumstellar disk \citep{gehrz1974infrared}. For CBeAe, the excess is predominantly concentrated in the NIR region. For HAeBe stars, the IR excess is attributed due to the thermal emission from the dust in the accretion disk \citep{1992ApJ...397..613H, 1998A&A...331..211M}.

In order to differentiate main-sequence ELS (CBeAe) and PMS stars (HAeBe), we used two proxies of IR excess. Firstly, we chose all the objects with Two Micron All-Sky Survey (2MASS; \citealp{2003yCat.2246....0C}) magnitudes satisfying the criteria that reddening corrected $(H-K_S)_0$ color is greater than 0.4, which is a well-established cut-off to select HAeBe stars \citep{1984A&AS...55..109F}. Secondly, we calculated two spectral indices \citep{LAda1987, Wilking1989, Green1994} from the SEDs as proxies to quantify NIR and mid-infrared (MIR) excess. NIR spectral index $(n_{J-K_S})$ is calculated using 2MASS $J$ and $K\textsubscript{S}$ magnitudes and MIR spectral index $(n_{K_S-W2})$, with 2MASS $K\textsubscript{S}$ and Wide-field Infrared Survey Explorer (WISE; \citealp{cutri2014vizier}) W2 magnitudes. The equation defining the spectral index is calculated as given in Equation \ref{eqn:3}. $\lambda_1$ and $\lambda_2$ represent the wavelength of two bands for which spectral index is calculated. $F_{\lambda_1}$ and $F_{\lambda_2}$ represent the extinction corrected absolute flux measured at $\lambda_1$ and $\lambda_2$.

\begin{figure}
    \centering
    \includegraphics[width=1.1\textwidth,height=1.44\textwidth, angle=0]{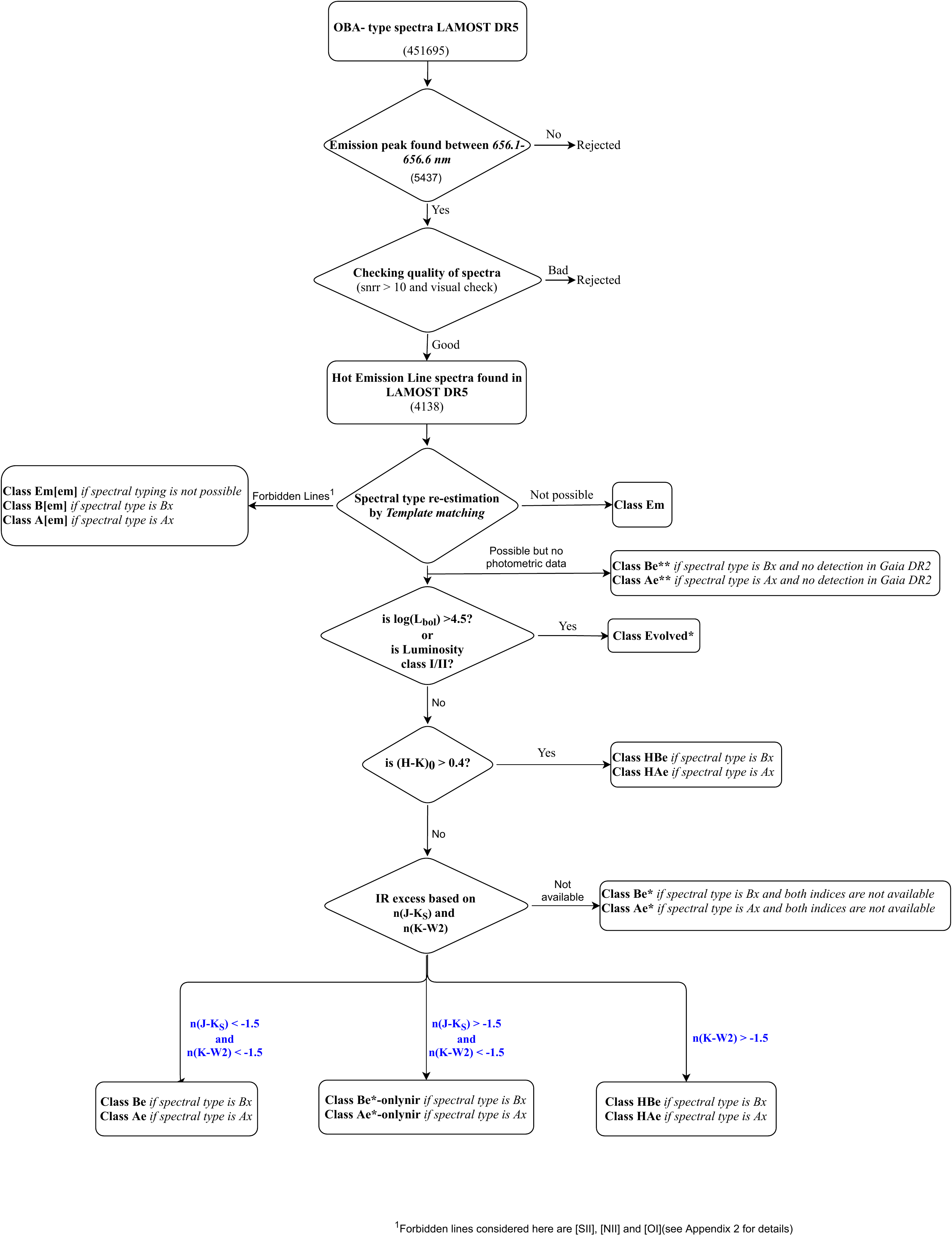}
    \caption{The Classification procedure used in this work is explained as flowchart here. Various classes that are used in each step are explained in detail.}
    \label{fig:flochart}
\end{figure}

\begin{equation}
n_{\lambda_1-\lambda_2} = \frac{log(\frac{\lambda_2F_{\lambda_2}}{\lambda_1F_{\lambda_1}})}{log(\frac{\lambda_2}{\lambda_1})}
\label{eqn:3}
\end{equation}

\begin{figure}
    \centering
    \includegraphics[width=\textwidth, angle=0]{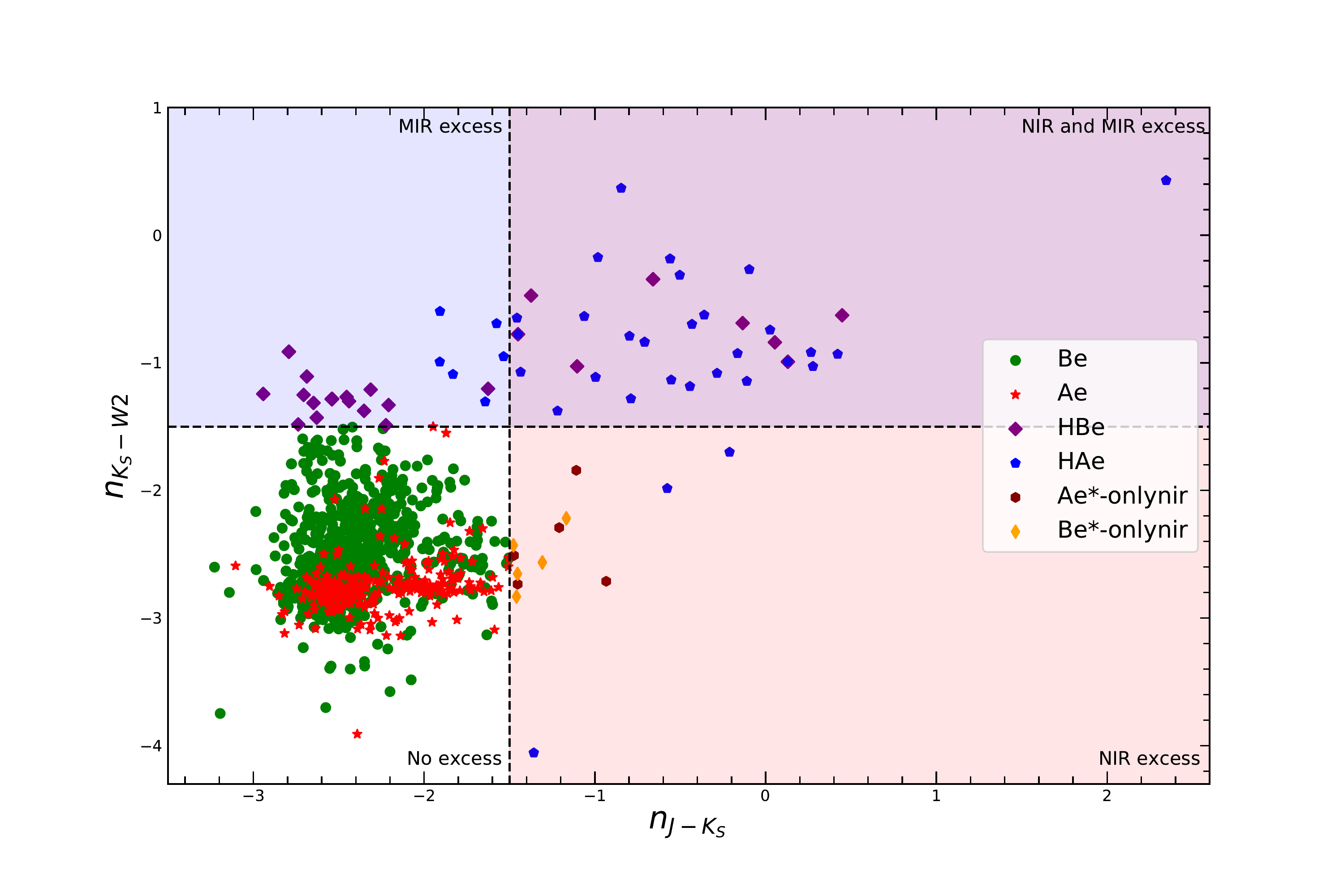}
    \caption{The plot illustrates classification of different ELS into categories based on their NIR and MIR Lada indices \protect\citep{LAda1987}. The black dashed lines represent $(n_{J-K_S})=$ -1.5 and $(n_{K_S-W2})=$ -1.5. Classes such as Be (green filled circles), Ae (red filled stars), HBe (purple filled diamonds), HAe (blue filled pentagons), Ae*-onlynir (brown filled hexagons), and Be*-onlynir (yellow solid thin diamonds) are shown in this figure. }
    \label{fig:index_plot}
\end{figure}

Based on the availability of IR photometric values for 3339 unique objects, $n_{J-K_S}$ was calculated for 2978 stars and $n_{K_S-W2}$ index was calculated for 2743 stars. Figure \ref{fig:index_plot} shows the classification of ELS based on $n_{J-K_S}$ and $n_{K_S-W2}$.  Remaining 597 stars did not have 2MASS or WISE magnitudes to calculate indices. We considered objects with $(n_{K_S-W2})>$ -1.5 to have MIR excess \citep{Andre1993, 2019yCat..51570159A}. In addition, we also segregated objects having $(n_{J-K_S})>$ -1.5 but $(n_{K_S-W2})<$ -1.5 into a separate class named ``Ae/Be*-onlynir". These objects have excess in NIR but do not show excess in MIR wavelengths, which needs to be studied in detail.

\subsection{Naming and Classification scheme}
\label{subsec:class}

 Even though we treated 4138 spectra as independent entities, many stars are observed multiple times during the 5-year observation cycle. Hence it is necessary to identify and name each spectra based on its observed coordinate. We sorted 4138 spectra based on its $snrr$ value in descending order and assigned name of first (highest SNR) spectra as ``LEMC 0001". LEMC is abbreviated as     ``LAMOST Emission-Line Catalog (LEMC)". If another spectrum has RA and DEC within 1\arcsec~ of ``LEMC 0001", then new spectrum is named as ``LEMC 0001$\_$2" and the next observation of same object (if present) is named as ``LEMC 0001$\_$3", and so on. The same process is iterated over the entire list and spectra are named from LEMC 0001 to LEMC 3339.

Coming to the classification of 4138 spectra, we created various classes to accommodate the non-availability of photometry values in some cases. If a spectra classified as B-type does not have \textit{Gaia} EDR3 or 2MASS photometry within search radius of 3\arcsec, then it is classified as ``Be**". Further if the same spectrum has \textit{Gaia} EDR3 and 2MASS photometry but was not detected in WISE, we classified it as ``Be*" since Lada spectral indices cannot be calculated. We classify a spectrum as ``CBe" only if it satisfies all three conditions, i.e., H$\alpha$ in emission, $(H-K_S)_0\leq$ 0.4 and the criteria set by Lada indices. This makes our classification robust.

We noted that in a few cases the multi-epoch observations have conflicting classifications. For example, LEMC 265 has three observations until DR5 which are classified as HBe, HAe and Em in each case. This is because spectra classified as ``Em" is noisy in the blue-end which made the spectral type estimation not possible. The spectral type of other two observations are estimated here to be B9 and A1V, which led to final classifications of HBe and HAe respectively. In another case, 7 multi-epoch observations of LEMC 0013 are classified into ``CBe" (4 spectra) and ``Evolved*" (3 spectra). Since the spectrum of LEMC 0013 with highest snrr values (LEMC 0013\_1) is classified as ``CBe", the star LEMC 0013 is put into list of 1089 unique CBe stars. Hence, care should be taken to select the best spectrum out of multiple observations.

\section{Results}
\label{sect:result}

We classified 4138 spectra of 3339 unique ELS identified from LAMOST DR5 into various classes based on available Optical/IR photometry and presence of forbidden emission lines. Detailed classification algorithm is shown in form of a flowchart in Figure \ref{fig:flochart}.

\begin{table}
\bc
\begin{minipage}[]{100mm}
\centering
\caption[]{Statistics of classified spectra\label{tab:statss}}\end{minipage}

\small
 \begin{tabular}{ccc|ccc}
  \hline\noalign{\smallskip}
Classification & Number of & Total number of & Classification & Number of & Total number of \\
 & unique objects & spectra available &  & unique objects & spectra available \\
  \hline\noalign{\smallskip}
CBe&1089&1523&CAe&233&286 \\
HAe&33&41&HBe&23&46 \\
A[em]&605&637&B[em]&323&391\\
Em&196&307&Em[em]&490&524 \\
Ae*&17&19&Be*&38&42 \\
Ae**&17&18&Be**&168&182 \\
Be*-onlynir&5&7&Ae*-onlynir&4&4 \\
Evolved*&36&46&F[em]?&58&61 \\
HFe?&1&1&Fe?*&3&3 \\

\hline
\end{tabular}
\ec
\end{table}


\subsection{2MASS-WISE Color-Color Diagram}

\begin{figure}
    \centering
    \includegraphics[width=1.0\textwidth, angle=0]{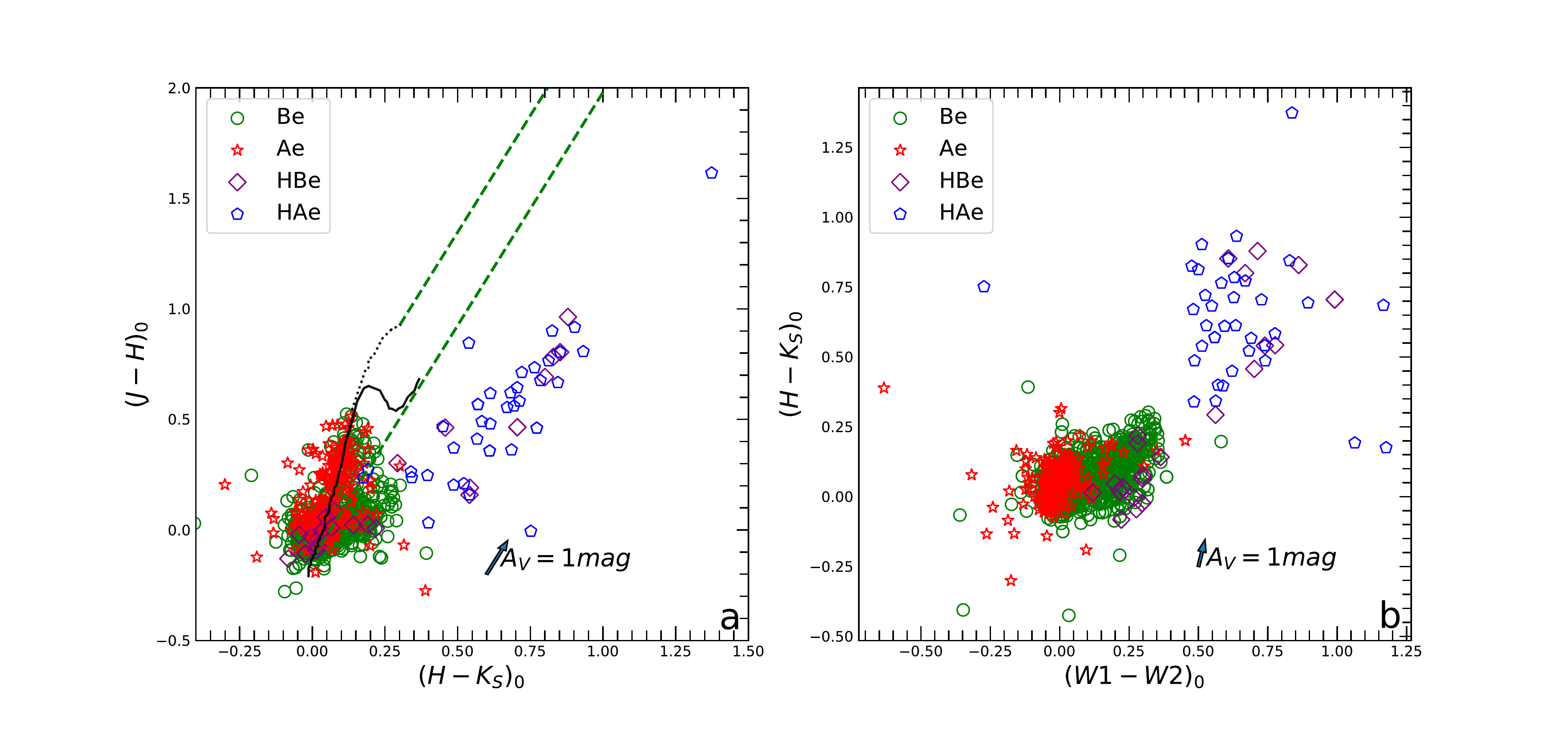}
    \caption{(a). Reddening corrected 2MASS CCDm showing positions of CBe (green open circles), CAe (red open stars), HBe (crimson red open squares) and HAe (blue open pentagons) stars. (b) Reddening corrected 2MASS-WISE CCDm showing positions of CBe (green open circles), CAe (red open stars), HBe (crimson red open squares) and HAe (blue open pentagons) stars.  The separation of main-sequence stars (CBe and CAe) from PMS stars (HBe and HAe) is very evident in the above plots. Arrow in both (a) and (b) represents the direction in which a star will move for a extinction of $A_{V}$ = 1 mag. }
    \label{fig:2MASS}
\end{figure}

Presence of dust in the circumstellar environment/accretion disk of the ELS can be identified from the excess flux emission in the infrared wavelengths \citep{2005ApJ...629..881H}. We plotted the $(H-K_S)_0$ versus $(J-H)_0$ for CBe, CAe and HAeBe stars in Figure \ref{fig:2MASS}. As observed in the 2MASS CCDm (see Figure \ref{fig:2MASS}), CBe stars are more clustered together than CAe stars. Also, note that most of the HAeBe stars have $(H-K_S)_0 >$ 0.4. It is seen that HAe stars strictly obey $(H-K_S)_0 >$ 0.4 cutoff than HBe stars as suggested in \cite{1984A&AS...55..109F}. IR excess in HBe stars is visibly evident in $(n_{K_S-W2})$ index than $(H-K_S)_0$ value.

Also, we have constructed a 2MASS-WISE CCDm where $(W1-W2)_0$ is plotted against $(H-K_S)_0$ and is shown in Figure \ref{fig:2MASS}. Almost all HAeBe stars occupy a distinct region compared to CBeAe stars, which justifies our classification technique.

\subsection{\textit{Gaia} Color-Magnitude Diagram}

\begin{figure}
    \includegraphics[width=1.08\textwidth,angle=0]{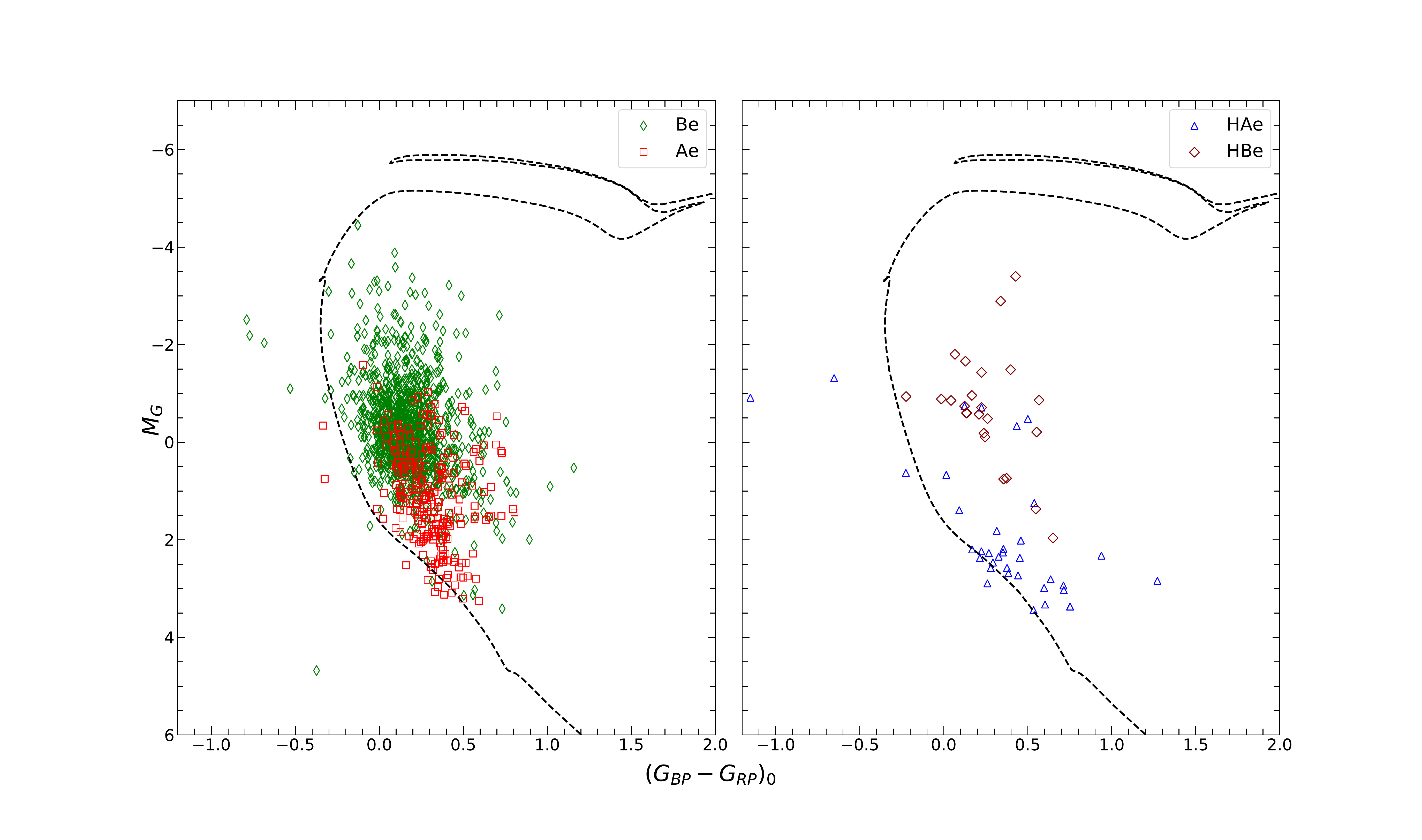}
    \caption{\textit{Gaia} CMD for CBeAe (left) and HAeBe stars (right) are shown. CBe (green open thin diamonds), CAe (red open squares), HBe (crimson red open diamonds) and HAe (blue open triangles) are shown. Dotted line in both the panels represents the isochrone of 60 Myr plotted in the \textit{Gaia} CMD with $(V/V_{crit} = 0.4)$ and $[Fe/H] = 0$.}
    \label{fig:optcmd}
\end{figure}

\textit{Gaia} EDR3 provides G, $G_{BP}$ and $G_{RP}$ magnitudes for 3017 ELS present in our sample. The absolute G magnitude ($M_G$) is calculated using the distance values taken from \citet{2021AJ....161..147B}. Further, the extinction corrected $M_G$ and $(G_{BP}-G_{RP})_0$ color are plotted in Figure \ref{fig:optcmd} for CBe, CAe and HAeBe stars. The Zero-age Main sequence (ZAMS) defined by \cite{2013ApJS..208....9P} does not extend beyond B9 spectral type for $M_G$ and $(G_{BP}-G_{RP})_0$ values.  Hence we plotted the 60 Myr isochrone from the Modules for Experiments in Stellar Astrophysics (MESA) isochrones and evolutionary tracks (MIST{\footnote{\url{http://waps.cfa.harvard.edu/MIST/} }}; \citealp{2016ApJ...823..102C,2016ApJS..222....8D}) in the \textit{Gaia} CMD (Figure \ref{fig:optcmd}), which closely matches the ZAMS defined by \cite{2013ApJS..208....9P}. The MIST archive has provided isochrone corresponding to $(V/V_{crit} = 0.4)$ which is the only model available in the database for a rotating system. We adopted the metallicity $[Fe/H] = 0$ for the respective isochrone. It is clear that the CAe and CBe stars occupy overlapping regions whereas HAe and HBe are well separated in \textit{Gaia} CMD.

\subsection{Spectral Features of various classes}
\label{subsec:stats}
In this subsection, we discuss the spectral features present in the stars in various evolutionary phases, as identified in this work. A breakup of 3339 ELS identified is shown in Table \ref{tab:statss}.
\begin{figure}
    \includegraphics[height=4.5in, width=6.5in, angle=0]{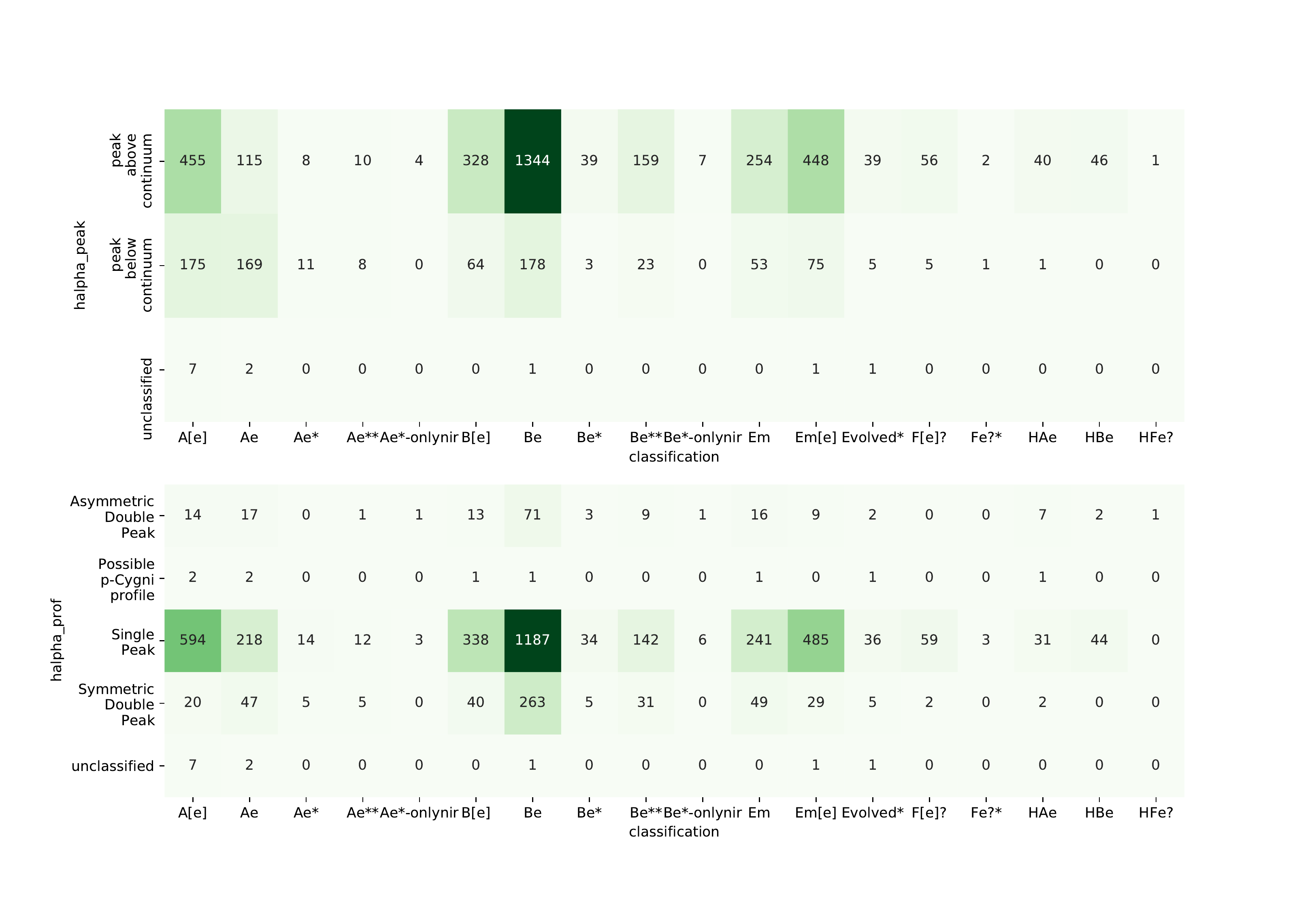}
    \caption{(top) A heatmap representing the statistical distribution of spectra based on classification of identified ELS spectra and position of H$\alpha$ peak in the spectra. (bottom) A heatmap representing the statistical distribution of spectra based on classification of identified ELS spectra and the morphology of H$\alpha$ peak in the spectra. The number of spectra belonging to each class is shown inside the respective boxes.}
    \label{fig:morph_crosstab}
\end{figure}

\subsubsection{Morphology of H$\alpha$}

In addition to classification of spectra explained in Section \ref{sect:Analysis}, we visually examined the spectra and noted different morphologies of H$\alpha$ emission seen in our sample. 
Out of 4138 ELS spectra identified in this work, 81\% of the spectra show strong emission with the H$\alpha$ peak above the continuum and remaining 19\% show weaker emission where the H$\alpha$ peak is seen inside the absorption core (emission-in-absorption). Further, 83\% of the spectra show single peak profile, 12\% show a symmetric double peak emission profile and 4\% show an asymmetric double peak emission profile. 
Statistical breakup of 4138 spectra based on H$\alpha$ profile morphology, H$\alpha$ peak and classification is shown in Figure \ref{fig:morph_crosstab} as a heatmap.

\subsubsection{Classical Be stars}

We identified a homogeneous sample of 1523 spectra belonging to 1089 CBe stars in this study. In addition to the photospheric absorption lines, the spectra of CBe stars generally show emission lines of different elements such as hydrogen, helium, oxygen, iron, calcium, etc. A detailed examination of these lines help us understand the properties of the gaseous disk.

As mentioned in previous literature \citep{2021MNRAS.500.3926B, 2013A&ARv..21...69R, doi:10.1016/j.nrjag.2012.12.004, mathew2011optical}, our sample of CBe stars also show diverse profiles in the Balmer lines. About 88\% of spectra classified as CBe show H{$\alpha$} in central emission above continuum and 12\% show emission-in-absorption. 78\% of spectra show single peak emission and 22\% of the spectra show double peak emission.  Around 45\% of the spectra show H{$\beta$} in weak emission-in-absorption. All of our sample stars show H{$\gamma$} and higher-order Balmer lines in absorption. Interestingly, we also observe variability in line profiles/intensity in some cases. Apart from hydrogen, we observe that few CBe stars in our sample also show O{\sc i} and Fe{\sc ii} lines in emission. Studies on the emission line features of FeII lines is sparse in literature when compared to H$\alpha$, due to the fact that Fe{\sc ii} lines are difficult to detect. Additionally, a closer view on the variability of Fe{\sc ii} lines compared to Balmer lines provide immense aid in understanding the kinematics of Be star disks. A dedicated analysis on these major line features observed in CBe stars will be presented as a follow-up work in Anusha et al. (in prep).

\subsubsection{Classical Ae stars}

We classified 286 spectra as CAe in this work with 169 spectra showing emission-in-absorption profile and 115 spectra showing emission above continuum. Since A-type stars have the strongest Balmer absorption lines, we observe more spectra with H$\alpha$ emission peak below the continuum. 76\% of the CAe spectra show single-peak emission, 16\%  show symmetric double-peak emission and only 6\% show asymmetric double-peak emission.

Out of 233 unique CAe stars identified, 159 CAe stars were analysed in detail by \cite{2021MNRAS.501.5927A}. The mismatch in number of CAe stars reported in their work and in our sample is due to a more stringent criteria of considering {$V$}-[12] magnitude color excess, where [12] is the IRAS 12 {$\mu$}m magnitude from IRAS Point Source catalogue and $V$ queried from APASS.
The CAe stars are expected to show weak H{$\alpha$} emission compared to CBe stars \citep{2013A&ARv..21...69R}. For the sample of 159 CAe stars, they observed that H{$\alpha$} EWs corrected for the underlying photospheric absorption are within the range of -0.2 to -23.6 {\AA}, which is comparatively lower than that reported for CBe stars. They also identified other emission lines such as  Fe{\sc ii}, O{\sc i}, Ca{\sc ii} triplet  and Paschen series in hydrogen lines from their sample of CAe stars.

\subsubsection{HAeBe stars}

We identified 87 spectra belonging to 56 HAeBe stars in this work. 86 out of 87 spectra show H$\alpha$ emission above the continuum. 87\% of HAeBe spectra show single-peak emission, 10\% show asymmetric double-peak emission and less than 2\% show symmetric double peak emission. It is interesting to see that 10\% of HAeBe show asymmetric double-peak emission whereas only 4\% of CBeAe spectra show asymmetric double-peak emission.

Apart from the presence of  Balmer lines, HAeBe stars share spectral similarities with those of CBeAe, which include O{\sc i} lines, Ca{\sc ii} infrared triplet and the Paschen series of H{\sc i} \citep{1985AJ.....90.1860P,1992ApJS...82..285H} are also seen in our sample of spectra. Other emission lines present in the spectra of HAeBe stars are Ca{\sc ii} H and K lines, Na{\sc i}, and Mg{\sc ii}. Since, we have removed the spectra with nebular forbidden lines from the analysis, the sample of HAeBe stars is not complete. Nidhi et al. (in prep) will consider this caveat and present an analysis of a bigger sample of intermediate-mass HAeBe stars from LAMOST DR5 including F0-F5 spectral types.

\begin{figure}
    \centering
    \includegraphics[width=1.1\textwidth]{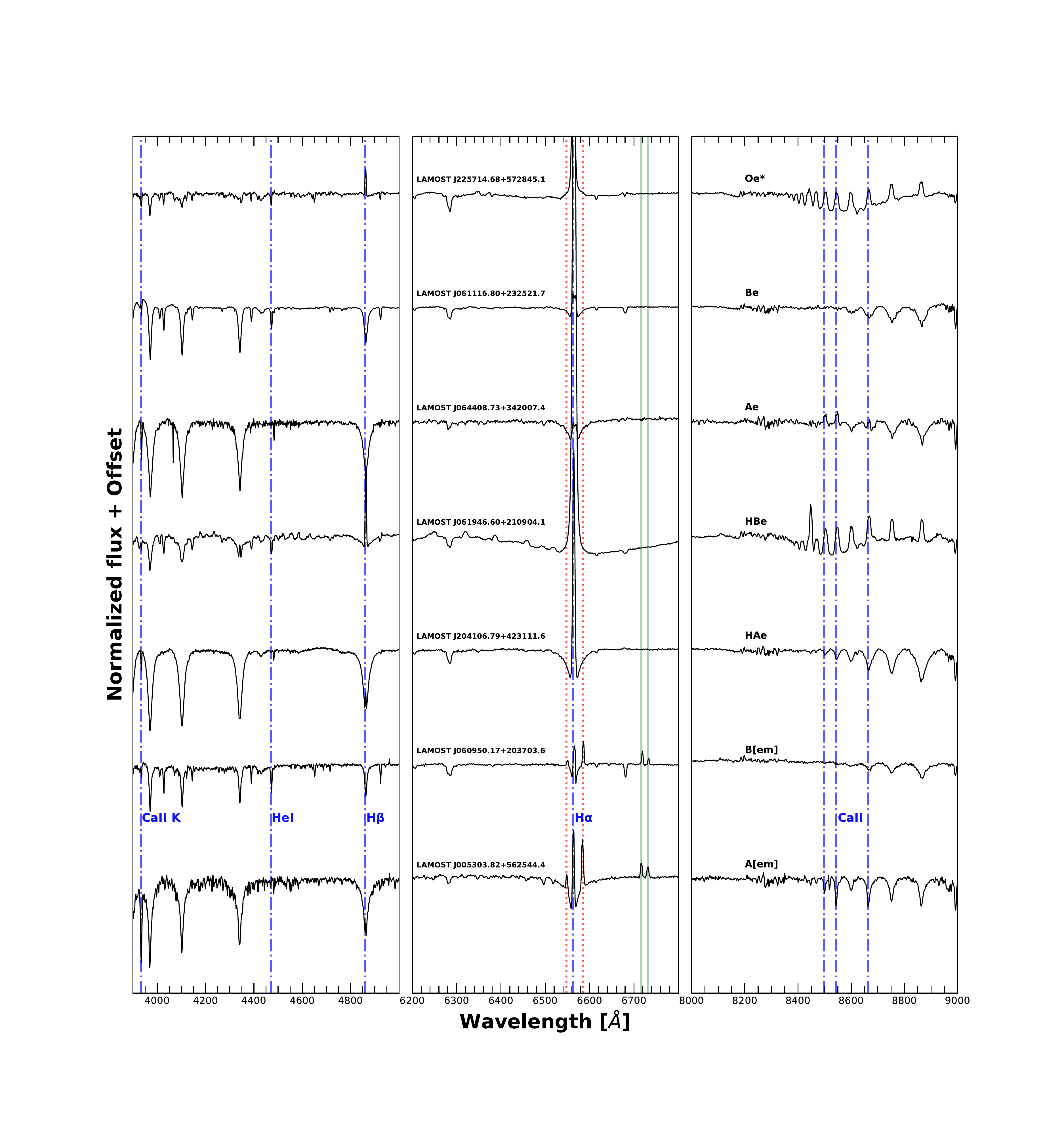}
    \caption{This plot shows a representative spectrum belonging to each of Oe*, CBe, CAe, HBe, HAe, B[em], and A[em] classes. Red dotted lines denotes [N{\sc{ii}}] lines at 6548 and 6584~\AA~and green solid lines denotes [S{\sc{ii}}] lines at 6717 and 6731~\AA. The spectra is normalized using \texttt{laspec} \citep{2020ApJS..246....9Z} module. }
    \label{fig:spec_plot}
\end{figure}
\subsection{SIMBAD crossmatch}
\label{subsec:simbad}

Finally, the newly identified ELS are cross matched with the SIMBAD database. Out of 3339 unique ELS, 623 objects have a record in SIMBAD database, among which 174 objects are marked as ``Em*" (Emission-line star) and 17 stars are recorded as ``Be*" (Be star) or ``Ae*" (Herbig Ae/Be star). 356 objects are represented as ``*" (Star) with no specific classifications. 22 objects are recorded as ``Y*O" (Young Stellar Object) or ``Y*?" (Young Stellar Object Candidate) and 7 objects are given as ``WD?" (White Dwarf candidate) by \citet{2013AJ....146...34Z}. Other types that are seen are ``IR" (IR source),``Ir*" (Variable Star of irregular type),``V*" (Variable star),``Ro*" (Rotationally Variable star) and ``HB*" (Horizontal Branch star). All of the objects marked in SIMBAD as ``Be*" are classified by us as ``Be" and objects marked as ``Y*O" are classified either as ``HAeBe" or ``A[em]/B[em]". Most of our classification is in accordance with what is reported in SIMBAD, the only notable differences are the 7 stars that are classified as ``WD?" by \citet{2013AJ....146...34Z}. In short, through this work of identifying the ELS from LAMOST DR5, we report 2716 new ELS which do not have any record on SIMBAD database. Figure \ref{fig:spec_plot} shows a representative sample of spectrum from some of the classifications discussed in this work.

\section{Conclusion}
\label{sect:conclusion}

We provide a catalog of 3339 hot (O, B and A spectral type) ELS identified from LAMOST DR5, of which 2716 are newly reported. Using an automated python routine, we identified 5437 spectra with H$\alpha$ in emission and visually selected 4138 good quality spectra for further analysis. We noticed that the spectral type provided by LAMOST is not accurate, deviating a few sub-classes in some cases. It is possible that emission features of the spectra is the reason for such mis-classifications. We re-estimated the spectral type of 3307 O, B and A type spectra by a semi-automated python routine that identifies three best template fits to the spectra in blue region including higher-order balmer lines such as H$\delta$ and H$\epsilon$ and then the best fit of the three templates (giving more importance to He{\sc{i}} and Mg{\sc{ii}} lines) was selected visually. Our re-estimated spectral type matches better with spectral type given in SIMBAD than spectral type provided by LASP, which validates our technique. 
Once the spectral types were re-estimated, we distinguished stars with excess in NIR and MIR bands based on 2MASS $(H-K_S)_0$, $(n_{J-K_S})$, and $(n_{K_{S}-W2})$ values. IR excess in NIR and MIR colors indicates the presence of dust in the circumstellar disk of PMS stars. CBe stars show low excess in NIR whereas HAeBe stars show considerable excess in NIR and sometimes a large excess in MIR as seen in Figure \ref{fig:2MASS}. We categorized the H$\alpha$ profiles of all 4138 emission-line spectra visually and it is represented as heatmap with respect to each classification, shown in Figure \ref{fig:morph_crosstab}.

We present in context a caveat associated with our classification. As explained in subsection \ref{subsec:class}, in some cases stars with multiple observations can end up having multiple classifications. This occurs due to the inability of distinguishing B8/9 type from A0/1 type spectra owing to low-resolution of LAMOST DR5 spectra. LAMOST Mid-Resolution Spectra (MRS) in DR6/7 will be helpful to resolve this confusion. 

 In this work we report a homogeneous list of 1089 CBe, 233 CAe, 56 HAeBe stars, and 686 stars categorized as ``Em" or ``Em[em]". In addition to these classes, we report 240 CBeAe candidates as ``Be*/**" and ``Ae*/**". It may be noted that 159 CAe stars identified from this work is analyzed in detail in \citet{2021MNRAS.501.5927A}. We classified 928 objects either as ``B[em]" or ``A[em]" due to presence of [S{\sc ii}], [N{\sc ii}] and/or [O{\sc i}] forbidden lines. Such a homogeneous list of emission-line catalog will help the community in studying ELS in detail without worrying about the bias associated with resolution and classification schemes when spectra is compiled from various sources. A sample of identified ELS is shown in Appendix \ref{appdx:C} and the entire catalog will be made available online in machine-readable format.

\begin{acknowledgements}
We would like to thank the Science \& Engineering Research Board (SERB), a statutory body of Department of Science \& Technology (DST), Government of India, for funding our research under grant number CRG/2019/005380. We thank the Center for Research, CHRIST (Deemed to be University), Bangalore, India, for funding our
research under the grant number MRP DSC-1932. We thank our colleagues Edwin Das and Robin Thomas for their valuable comments on the manuscript. This work has made use of data products from the Guo Shoujing Telescope (the Large Sky Area Multi-Object  Fibre  Spectroscopic  Telescope,  LAMOST). This work has made use of data from the European Space Agency (ESA) mission
{\it Gaia} (\url{https://www.cosmos.esa.int/gaia}), processed by the {\it Gaia}
Data Processing and Analysis Consortium (DPAC, \url{https://www.cosmos.esa.int/web/gaia/dpac/consortium}). Funding for the DPAC has been provided by national institutions, in particular the institutions
participating in the {\it Gaia} Multilateral Agreement. We thank the SIMBAD database and the online VizieR library service for helping us in literature survey and obtaining relevant data.
\end{acknowledgements}

\appendix                  

\newpage

\section{Catalog Description}
\label{appdx:A}
\begin{table}[!h]
\caption{Description of columns present in the catalog}
\centering
\begin{tabular}{| p{5.50cm} | m{10.50cm}|l|c|}
\toprule
{\textbf{Columns}} & \textbf{Description} \\ 
\midrule
$specname$ & Name of FITS file as given by LAMOST \\
\hline
$classification$ & Class that the spectra belongs to \\
\hline
$catalog\_name$, $unique\_name$  & Name assigned to the spectra in this work. $unique\_name$ gives the name of ELS star the spectra belongs. In addition to $unique\_name$, $catalog\_name$ contains an extra number to identify spectra with highest $snrr$ value. That is, if an ELS has multi-epoch spectra, then the highest $snrr$ spectra will be denoted by ``$\_1$" in its $catalog\_name$ \\
\hline
$spectral\_type$ & Spectral type assigned to the spectra. The spectral type of best fit template from MILES library is provided. \\
\hline
$halpha\_profile\_comments$ & Comments on H-alpha emission profile, visually checked \\
\hline
$forb\_visual\_comments$ & Comments on forbidden lines present in spectra, visually checked \\

\hline
$l6300$, $l6363$, $l6548$, $l6584$, $l6717$, $l6731$ & ``yes" or ``no", based on an automated line finder routine. The code and threshold used is explained in Appendix 2. \\
\hline
$lamost\_design\_id$ & The ID as provided in LAMOST DR5 \\
\hline
$\_RAJ2000$, $\_DEJ2000$ & Observed RA and Dec in degrees as given in FITS header \\
\hline
$subclass$ & Spectral type assigned by LASP, as given in LAMOST DR5. \\
\hline
$snru$, $snrg$, $snrr$, $snri$, $snrz$ & Signal-to-Noise Ratio of the spectrum in different SDSS bands, as given in LAMOST DR5. \\

\hline
$2MASS$~to~$Rfl$ & 2MASS photometry \\
\hline
$ALLWISE$~to~$qph$ & ALLWISE photometry \\
\hline
$B-V$~to~$u\_e\_Bmag$ & APASS photometry \\
\hline
$Av$, $Av\_lower$, $Av\_upper$ & Extinction in V band queried from \citet{2019ApJ...887...93G}. Upper and Lower bounds are given based on $r\_low\_photogeo$ and $r\_hi\_photogeo$ respectively \\
\hline
$n(J-K)$, $n(K-W2)$ & Lada indices calculated as defined in sect. \ref{subsec:sed} \\
\hline
$(J-H)_0$~to~$e\_(H-K)_0$ &  Extinction corrected 2MASS and WISE colors\\
\hline
$abs\_Vmag$ to $(B-V)_0$ & Extinction corrected APASS absolute V magnitude along with errors and (B-V) color\\
\hline
$gaiaedr3\_source\_id$~to
\newline~$phot\_rp\_mean\_mag\_error$ & $Gaia$ EDR3 photometry\\
\hline
$r\_med\_geo$ to $r\_hi\_photogeo$ & Distances (geometric and photo-geometric) to the star along with upper and lower bounds based on $Gaia$ EDR3 taken from \citet{2021AJ....161..147B}. ``geometric" distance is calculated using parallax and a direction dependent distance prior, whereas ``photo-geometric" used an additional color and apparent magnitude of the star. \\
\bottomrule
\end{tabular}

\label{tab:catdescp}
\end{table}

\section{Forbidden Lines}
\label{appdx:B}

To distinguish forbidden lines from noisy features in low-SNR spectra, we used the `prominence' feature in `find\_peaks' module.
We flagged a forbidden line as ``yes" only if the prominence is greater than 0.2; if prominence value is between 0.1 and 0.2, it is flagged as ``maybe" since it can be difficult to distinguish it from a noise feature if SNR is low and if prominence value is less than 0.1, the line is flagged as ``no". As CBeAe stars do not show forbidden lines such as [S{\sc ii}] $\lambda\lambda$ 6717,6731, [N{\sc ii}] $\lambda\lambda$ 6548,6584 and [O{\sc i}] $\lambda\lambda$ 6300,6363 in their spectra, it is necessary to identify and remove spectra with above mentioned forbidden lines. 
 
In addition to each forbidden line's flag, the final decision was taken by visually selecting the spectra without any forbidden lines. We included the flags keeping in mind the future works, where one can use the flags given to select a subset of spectra with specific forbidden lines.

\section{Sample of ELS identified}
\label{appdx:C}

\begin{longtable}{|cccccc|}
\caption{Details of sample of ELS identified in our work}\\
\hline
lamost$\_$design$\_$id&$\_$RAJ2000&$\_$DEJ2000&classification&spectral$\_$type&catalog$\_$name \\

\midrule

LAMOST J043018.87+071233.5&67.578651&7.209331&A[em]&A1V&LEMC 2910 \\
LAMOST J043020.98+455603.2&67.58745&45.934232&Be&B8III&LEMC 0675 \\
LAMOST J043023.15+550408.8&67.596482&55.069113&Be*-onlynir&B8III&LEMC 0495$\_$1 \\
LAMOST J043025.78+490355.8&67.607458&49.0655&Em&&LEMC 2122 \\
LAMOST J043030.15+533334.7&67.625642&53.559666&Be&B5V&LEMC 0493 \\
LAMOST J043041.64+432544.5&67.673508&43.429055&Be&B5V&LEMC 1863 \\
LAMOST J043058.53+373855.7&67.743908&37.64883&Evolved*&A2Ib/II&LEMC 0458$\_$1 \\
LAMOST J043104.88+520145.7&67.770358&52.029388&Em[em]&&LEMC 1862 \\
LAMOST J043111.79+322932.7&67.79914&32.492425&Be&B9III&LEMC 0896$\_$1 \\
LAMOST J043131.26+475750.7&67.880284&47.964108&Be&B2IIIvar&LEMC 0204$\_$1 \\
LAMOST J043133.92+523039.2&67.891338&52.510914&A[em]&A0III&LEMC 3203 \\
LAMOST J043134.99+493650.1&67.895833&49.613944&Em&&LEMC 2975 \\
LAMOST J043153.91+430248.7&67.974655&43.046862&Be&B6IV&LEMC 2246 \\
LAMOST J043220.04+513458.1&68.083504&51.582813&Ae&A1V&LEMC 2131 \\
LAMOST J043226.98+343056.2&68.11243&34.515624&B[em]&B8V&LEMC 0701 \\
LAMOST J043229.03+540352.2&68.120962&54.064511&Be&B8III-IV&LEMC 0408$\_$1 \\
LAMOST J043230.94+531036.0&68.128957&53.176687&Be&B5V&LEMC 1774$\_$1 \\
LAMOST J043244.07+554120.0&68.183645&55.688909&Be&B8&LEMC 0095$\_$1 \\
LAMOST J043247.86+455849.2&68.199448&45.980349&B[em]&B8V&LEMC 1947 \\
LAMOST J043308.59+382811.8&68.285831&38.469964&Em[em]&&LEMC 1628$\_$1 \\
LAMOST J043324.72+405849.1&68.353&40.980333&Be&B8&LEMC 1382$\_$1 \\
LAMOST J043327.11+512242.3&68.362983&51.37842&Be**&B8&LEMC 0020 \\
LAMOST J043327.49+512242.4&68.364582&51.378454&Be**&B8&LEMC 0909 \\
LAMOST J043347.36+330254.0&68.44737&33.048357&Em[em]&&LEMC 2725 \\
LAMOST J043431.86+442815.7&68.632762&44.471051&Be&B8V&LEMC 1205 \\
LAMOST J043434.43+505322.9&68.643483&50.88972&Be&B9p+...&LEMC 1663$\_$1 \\
LAMOST J043434.61-025615.8&68.644226&-2.937746&Ae*&A1IV&LEMC 2382 \\

\hline
\label{tab:tabb}
\end{longtable}

\bibliographystyle{mnras}
\bibliography{bibtex}

\label{lastpage}

\end{document}